\documentclass[aps,pra,twocolumn,showpacs,floatfix]{revtex4}
\usepackage{epsfig}
\usepackage{graphicx}
\usepackage{dcolumn}
\usepackage{amsthm,amsmath}

\begin{document}
\title{State-insensitive trapping of Rb atoms: linearly versus circularly polarized lights}
\author{Bindiya Arora} 
\affiliation{Department of Physics, Guru Nanak Dev University, Amritsar, Punjab-143005, India} 

\author{B. K. Sahoo \footnote{Email: bijaya@prl.res.in}}
\affiliation{Theoretical Physics Division, Physical Research Laboratory, Ahmedabad-380009, India}
\date{Received date; Accepted date}
 
\begin{abstract}
We study the cancellation of differential ac Stark shifts in the 
$5s$ and $5p$ states of rubidium atom using the linearly and circularly 
polarized lights by calculating their dynamic polarizabilities. Matrix
elements were calculated using a relativistic coupled-cluster method at the single,
double and important valence triple excitations approximation including all possible
non-linear correlation terms. Some of the important matrix elements were
further optimized using the experimental results available for the lifetimes and 
static polarizabilities of atomic states. ``Magic wavelengths" 
are determined from the differential Stark shifts and results for the linearly 
polarized light are compared with the previously available results. Possible
scope of facilitating state-insensitive optical trapping schemes using the
magic wavelengths for circularly polarized light are discussed. Using the 
optimized matrix elements, the lifetimes of the $4d$ and $6s$ states of
this atom are ameliorated.
\end{abstract}

\maketitle

\section{Introduction}\label{sec1}
Investigating the properties of rubidium (Rb) atom is of immense interest for a number of applications
\cite{Rb1,Rb2,Rb3,Rb4,Rb5,Rb6,Rb7,Rb8,Rb9,pnc,sahoo1}. It is one of the most widely used atom 
in quantum computational schemes 
using Rydberg atoms, where the hyperfine states of the ground state of Rb atom 
are defined as the qubits \cite{Rb4}. It is also used to study quantum phase transitions
of mixed-species with degenerate quantum gases \cite{Rb6}. There are several proposals
to carry out precision studies in this atom such as constructing ultra-precise atomic
clocks \cite{Rb7,Rb8,Rb9}, probing parity non-conservation effects \cite{pnc}, finding 
its permanent electric dipole moment \cite{sahoo1} etc. Also, a number of measurements
and calculations of lifetimes for many low-lying states in Rb have been performed 
over the past few decades \cite{life1,life2,life3,life4,life5,life6,life7}. It is found that there are inconsistencies 
between the calculated and measured values of the lifetimes of atomic states in this atom
\cite{life7}. In this context, it is necessary to carry out further theoretical
studies in this atom. 

Due to simple single-core electron structure of this atom, it is adequate to employ advanced
many-body methods for precise calculation of its properties which ultimately act as benchmark tests for the experimental measurements \cite{ben1,ben2,ben3}.
In this paper, we determine polarizabilities of the ground $5s$ and excited $5p$ states and study the
differential ac Stark shifts between these two states. In this process, we also analyse
the reduced matrix elements and their accuracies which are further used to estimate precisely the 
lifetimes of few excited states in this atom. Aim of our present study is to analyse results
of differential ac Stark shifts from which we can deduce the magic wavelengths (see below
for definition) that are of great use in state-insensitive trapping of Rb atoms.

Manipulation of cold and ultracold Rb atoms has been widely done by using optical traps
\cite{dipole1,dipole2}. For a number of applications (such as atomic clocks and quantum 
computing~\cite{AC1,QC1}), it is often desirable to optically trap the neutral atoms
without affecting the internal energy-level spacing for the atoms. However in an 
experimental set up, the interaction of an atom with the externally applied oscillating 
electric field of the trapping beam causes ac Stark shifts of the atomic levels inevitably. 
For any two internal states of an atom, the Stark shifts caused due to the trap
light are in general different which affects the fidelity of the experiments \cite{QC2,katori1}.
Katori \textit{et al.} \cite{katori2} proposed the idea of tuning the trapping laser to a 
magic wavelength, ``$\lambda_{\rm{magic}}$", at which the differential ac Stark shifts of the transition 
is terminated. Using this approach the magic wavelength for the $5s^2 \ ^1S_0^0 \ - \ 5s5p \ ^3P_0^0$ transition in $^{87}$Sr
was determined with a high precision to be 813.42735(40) $nm$ \cite{ludlow}. McKeever \textit{et al.} 
demonstrated the state-insensitive trapping of Cs
at $\lambda_{\rm{magic}} \approx$ 935 $nm$ while still maintaining a strong coupling with the
$6s_{1/2}-6p_{3/2}$ transition \cite{kimble1}. 
Arora \textit{et al.} \cite{arora1} calculated the magic wavelengths for the $np-ns$ transitions 
for other alkali atoms (from Na to Cs) by calculating dynamic 
polarizabilities using a relativistic coupled-cluster (RCC) method.
Theoretical values for these quantities were calculated at wavelengths where 
the ac polarizabilities for two states involved in the transition cancel. The data in Ref. 
\cite{arora1} provides a wide range of magic wavelengths for the alkali-metal atoms trapped 
in linearly polarized light by evaluating electric dipole (E1) matrix elements obtained by linearized
RCC method. In this paper, we try to evaluate these matrix elements considering all possible
non-linear terms in the RCC method. In addition, we would like to optimize the matrix elements
using the precisely known experimental results of lifetimes and
static polarizabilities for different atomic states and re-investigate the above reported magic 
wavelengths in the considered atom. It is also reported in Ref. \cite{arora1} that trapping
Rb atoms in the linearly polarized light offers only a few suitable magic wavelengths for  
the state-insensitive scheme. This persuades us to look for more plausible cases for 
constructing state-insensitive traps of Rb atoms using the circularly polarized light.
Using the circularly polarized light may be advantageous owing to the 
dominant role played by vector polarizabilities (which are absent in the linearly polarized
light) in estimating the ac Stark shifts. Moreover, these vector polarizabilities
 act as ``fictitious magnetic fields", turning the ac Stark shifts to the case 
analogous to the Zeeman shifts \cite{derevianko1,zeeman1}.

This paper is organized as follows, in sections \ref{sec2} and \ref{sec3}, we discuss 
in brief the theory of dipole polarizability and method used for calculating them precisely. In section \ref{sec4}, we first discuss in 
detail the evaluation of matrix elements used for precise estimation of 
polarizability and then present our magic wavelengths first for the linearly 
polarized light following which for the circularly polarized light.
Unless stated otherwise, we use the conventional system of atomic units (au), in which $e$, $m_e$, 4$\pi \epsilon_0$, and the reduced Planck constant $\hbar$ have the numerical value 1 throughout this paper.

\section{Theory of dipole polarizability}\label{sec2}
The $v^{\rm{th}}$ energy level of an atom placed in a static electric
field $\mathcal{E}$ can be expressed using a time-independent perturbation
theory as \cite{pol-book}
\begin{equation}
E_v={E_v}^0+\sum_{k\neq v}\frac{| \langle \psi_v\left|V\right|\psi_k \rangle |^2}{{E_v}^0-{E_k}^0}+\cdots,
\end{equation}	
where ${E_i}^0$s are the unperturbed energy levels in the absence of electric
field, $k$ represent the intermediate states allowed by the dipole selection rules and $V=-{D}\cdot\mathcal{E}$ is the interaction Hamiltonian with ${D}$ as 
the electric-dipole operator. Since the 
first-order correction to the energy levels is zero
in the present case, therefore we can approximate the energy shift at the second-order
level for a weak field $\mathcal{E}$ and write it in terms of dipole moments 
$p$ as
\begin{equation}
\Delta E_v = E_v-{E_v}^0 \simeq \sum_{k \neq v}\frac{({p^*})_{vk}({p})_{kv}}{\delta E_{vk}}\mathcal{E}^2,\label{ch1comp1}
\end{equation}	
where $\delta E_{vk}=({E_v}^0-{E_k}^0)$ and $(p)_{vk}=\left\langle \psi_v\left|D\right|\psi_k\right\rangle$ is the E1 amplitude between $|\psi_v\rangle$ and $|\psi_k\rangle$ states.
A more traditional notation of the above equation is given by
\begin{equation}
\Delta E_v=-\frac{1}{2}\alpha_v \mathcal{E}^2, \label{eq-1}
\end{equation}	
where $\alpha_v$ is known as the static polarizability of the $v^{th}$ state
which is written as
\begin{eqnarray}
\alpha_v = - 2 \sum_{k \neq v}\frac{({p^*})_{vk}({p})_{kv}}{\delta E_{vk}}.
\end{eqnarray}
If the applied field is frequency-dependent (ac field), then we can still
express the change in energy as $\Delta E_v=-\frac{1}{2}\alpha_v \mathcal{E}^2$
with $\alpha_v$ as a function of frequency given by
\begin{eqnarray}
\alpha_v (\omega) &=& - \sum_{k \neq v} ({p^*})_{vk}({p})_{kv} \left [ \frac {1}{\delta E_{vk} + \omega } + \frac {1}{\delta E_{vk} - \omega } \right ]. \ \ \ \
\end{eqnarray}

Since $\alpha_v (\omega)$ also depends on angular momentum $j$ and $m_j$ values of the given
atomic state, it is customary to express them in a different form with
$m_j$ dependent factors and $m_j$ independent factors. Therefore, $\alpha_v (\omega)$
is further rewritten as \cite{manakov}
\begin{eqnarray}
\alpha_v (\omega) &=& \alpha_v^0(\omega) + \mathcal{A} \cos{\theta_k} \frac{m_j}{j} \alpha_v^1(\omega) \nonumber \\ && + \left \{ \frac{3 \cos^2{\theta_p} -1}{2} \right \} \frac{3m_j^2 - j(j+1)}{j(2j-1)} \alpha_v^2(\omega),
\label{cpl}
\end{eqnarray}
where $\mathcal{A}$, $\theta_k$ and $\theta_p$ define degree of circular polarization, 
angle between wave vector of the electric field and $z$-axis and angle
between the direction of polarization and $z$-axis, respectively. Here $\mathcal{A}=0$ for the 
linearly polarized light implying there is no vector component present in this case; otherwise 
$\mathcal{A}=1$ for the right-handed
and $\mathcal{A}=-1$ for the left-handed circularly polarized light. In the absence of
magnetic field (or in weak magnetic field), we can choose $\cos(\theta_k ) = \cos( \theta_p ) = 1$. 
Here $m_j$ independent factors $\alpha_v^0$,
$\alpha_v^1$ and $ \alpha_v^2$ are known as scalar, vector and tensor 
polarizabilities, respectively. In terms of the reduced matrix elements of 
dipole operator they are given by \cite{manakov}
\begin{eqnarray}
\alpha_v^{0}(\omega) &=& \frac{1}{{3(2j_v+1)} }\sum_{j_k}
|\langle \psi_v \parallel D\parallel \psi_k \rangle|^{2} \nonumber \\ && \times \left [ \frac{1}{\delta E_{kv}+\omega}+\frac{1}{\delta E_{kv}-\omega} \right ],\label{eq-scalar} \\
\alpha_v^{1}(\omega) &=& - \sqrt { \frac{6j_v} {(j_v+1)(2j_v+1)} }
\sum_{j_k} \left \{ \begin{array}{ccc} j_v & 1 & j_v \\ 1 & j_k & 1 \end{array} \right \} \nonumber \\ && (-1)^{j_v+j_{k}+1} 
 | \langle \psi_v \parallel D \parallel \psi_k \rangle|^{2} \nonumber \\ && \times \left [ \frac{1}{\delta E_{kv}+\omega}-\frac{1}{\delta E_{kv}-\omega} \right ] \label{eq-vector} \\
\alpha_v^{2}(\omega) &=& -2 \sqrt { \frac{5j_v(2j_v-1)} {6(j_v+1)(2j_v+1)(2j_v+3)} } \nonumber \\ && 
\sum_{j_k} \left \{ \begin{array}{ccc} j_v & 2 & j_v \\ 1 & j_k & 1 \end{array} \right \} (-1)^{j_v+j_{k}+1} 
 | \langle \psi_v \parallel D \parallel \psi_k \rangle|^{2}  \nonumber \\ && \times \left [ \frac{1}{\delta E_{kv}+\omega}+\frac{1}{\delta E_{kv}-\omega} \right ]\label{eq-tensor}.
\end{eqnarray}
For $\omega=0$, the results will correspond to the static polarizabilities 
which clearly suggests that $\alpha_v^1$ is zero for the static case.

\section{Method of calculations}\label{sec3}
To calculate wave functions in Rb atom, we first obtain Dirac-Fock (DF) wave function
for the closed-shell configuration $[4p^6]$ which is given by $|\Phi_0\rangle$. Then the DF
wave function for atomic states with one valence configuration are defined as 
\begin{eqnarray}
| \Phi_v \rangle =a_v^{\dagger}|\Phi_0\rangle,
\end{eqnarray}
where $a_v^{\dagger}$ represents addition of the valence orbital, denoted by $v$, with 
$|\Phi_0\rangle$. The exact atomic wave function ($|\Psi_v \rangle$) for such a configuration 
is determined, accounting correlation effects in the RCC framework, by expressing \cite{rcc}
\begin{eqnarray}
|\Psi_v \rangle &=& e^T \{1+S_v\} |\Phi_v \rangle ,
\label{cc1}
\end{eqnarray}
which in linear form is given by
\begin{eqnarray}
|\Psi_v \rangle & \approx & \{1+T+S_v\} |\Phi_v \rangle .
\label{cc2}
\end{eqnarray}
Here $T$ and $S_v$ operators account excitations of the electrons from the
core orbitals alone and valence orbital together with core orbitals, respectively.
In the present paper, we consider Eq. (\ref{cc1}) instead of Eq. (\ref{cc2})
as was taken before in our previous calculations \cite{arora1}. We consider here only
single, double (CCSD method) and important triple excitations (known as
CCSD(T) method from $|\Phi_0\rangle$ and $| \Phi_v \rangle$. 

The excitation amplitudes for the $T$ operators are determined by solving
\begin{eqnarray}
\langle \Phi_0^* |\{\widehat{He^T}\}|\Phi_0 \rangle &=& 0, 
\label{eqn23}
\end{eqnarray}
where $|\Phi_0^* \rangle$ represents singly and doubly excited configurations from 
$| \Phi_0 \rangle$. Similarly, the excitation amplitudes for the $S_v$ operators are 
determined by solving
\begin{eqnarray}
\langle \Phi_v^*|\{\widehat{He^T}\} \{1+S_v\}|\Phi_v\rangle &=& \langle \Phi_v^*|S_v|\Phi_v\rangle \Delta E_v^{att},
\label{eqn24}
\end{eqnarray}
taking $|\Phi_v^* \rangle$ as the singly and doubly excited configurations from $| \Phi_v \rangle$. The
above equation is solved simultaneously with the calculation of attachment energy $\Delta E_v^{att}$
for the valence electron $v$ using the expression
\begin{eqnarray}
\Delta E_v^{att} = \langle \Phi_v|\{\widehat{He^T}\} \{1+S_v\} |\Phi_v\rangle.
\end{eqnarray}
The triples effect are incorporated through the calculation of $\Delta E_v^{att}$ by including
valence triple excitation amplitudes perturbatively (e.g. see \cite{bks} for the detailed discussion).

To determine polarizabilities, we divide various correlation contributions to it into
three parts as
\begin{eqnarray}
\alpha_v^{\lambda}  &=& \alpha_v^{\lambda}(c) + \alpha_v^{\lambda}(vc) + \alpha_v^{\lambda}(v),
\end{eqnarray}
where $\lambda=0, \ 1$ and $2$ represents scalar, vector and tensor
polarizabilities, respectively, and the notations $c$, $vc$ and $v$ in the
parentheses correspond to core, core-valence and valence correlations,
respectively. The core contributions to vector and tensor polarizabilities
are zero.

 We determine the valence correlation contributions to the polarizability in the
sum-over-states approach \cite{arora-sahoo1} by evaluating their matrix elements 
by our CCSD(T) method and using the experimental energies \cite{NIST,NIST1,NIST2} 
for the important intermediate states. Contributions from the
higher excited states and continuum are accounted from the following
expression
\begin{eqnarray}
\alpha_v^{\lambda} &=& C_{\lambda} \langle \Psi_v | D | \Psi_v^{(1)} \rangle,
\end{eqnarray}
where $C_{\lambda}$ are the corresponding angular factors for different values
of $\lambda$ and $| \Psi_v^{(1)} \rangle$ is treated as the first order wave 
function to $| \Psi_v \rangle$ due to the dipole operator $D$ \cite{sahoo4} at the third
order many-body perturbation (MBPT(3) method) level and given as
$\alpha_v^{\lambda}(\rm{tail})$. Also, contributions from the core and core-valence
correlations are estimated using this procedure.

We calculate the reduced matrix elements of $D$ between states $| \Psi_f \rangle$ and 
$| \Psi_i \rangle$, to be used in the sum-over-states approach, from the following RCC
expression
\begin{eqnarray}
\langle \Psi_f || D || \Psi_i \rangle &=& \frac{\langle \Phi_f || \{ 1+ S_f^{\dagger}\} \overline{D } \{ 1+ S_i\} ||\Phi_i\rangle}{ \sqrt{{\cal N}_f {\cal N}_i}}, 
\end{eqnarray}
where $\overline{ D}=e^{T^{\dagger}} D e^T$ and ${\cal N}_v = \langle \Phi_v |
e^{T^{\dagger}} e^T + S_v^{\dagger} e^{T^{\dagger}} e^T S_v |\Phi_v\rangle$ 
involve two non-truncating series in the above expression. Calculation 
procedures of these expressions are discussed in detail elsewhere 
\cite{mukherjee,sahoo2}.

\section{Results and Discussion}\label{sec4}

Our aim is to determine the magic wavelengths for the linearly and circularly polarized 
electric fields for the $5s-5p_{1/2,3/2}$ transitions in Rb atom. To determine these
wavelengths precisely, we need accurate values of polarizabilities which depend upon the
excitation energies and the E1 matrix elements between the intermediate states of the
corresponding states. In this respect, we first
present below the E1 matrix elements between different transitions and discuss 
their accuracies. Then we overview the current status of
the polarizabilities reported in literature and compare our results with
them. These results are further used to determine the magic wavelengths for both
the linearly and circularly polarized lights.

\begin{table}
\caption{\label{e1mat} Absolute values of E1 matrix elements in Rb atom in $ea_0$ using the Dirac-Fock (DF) and CCSD(T) methods. Uncertainties in the CCSD(T) results are given in the parentheses.}
\begin{ruledtabular}
\begin{tabular}{lcc}
Transition & DF & CCSD(T) \\
\hline
$5s_{1/2} \rightarrow 5p_{1/2}$  & 4.819 &  4.26(3)  \\ 
$5s_{1/2} \rightarrow 6p_{1/2}$  & 0.382 &  0.342(2) \\
$5s_{1/2} \rightarrow 7p_{1/2}$  & 0.142 &  0.118(1) \\
$5s_{1/2} \rightarrow 8p_{1/2}$  & 0.078 &  0.061(5) \\
$5s_{1/2} \rightarrow 9p_{1/2}$  & 0.052 &  0.046(3)  \\
$5s_{1/2} \rightarrow 5p_{3/2}$  & 6.802 &  6.02(5)  \\
$5s_{1/2} \rightarrow 6p_{3/2}$  & 0.605 &  0.553(3) \\
$5s_{1/2} \rightarrow 7p_{3/2}$  & 0.237 &  0.207(2) \\
$5s_{1/2} \rightarrow 8p_{3/2}$  & 0.135 &  0.114(2) \\
$5s_{1/2} \rightarrow 9p_{3/2}$  & 0.091 &  0.074(2) \\
$5p_{1/2} \rightarrow 6s_{1/2}$  & 4.256 &  4.144(3) \\
$5p_{1/2} \rightarrow 7s_{1/2}$  & 0.981 &  0.962(4) \\
$5p_{1/2} \rightarrow 8s_{1/2}$  & 0.514 &  0.507(3) \\
$5p_{1/2} \rightarrow 9s_{1/2}$  & 0.337 &  0.333(1) \\
$5p_{1/2} \rightarrow 10s_{1/2}$ & 0.239 &  0.235(1) \\
$5p_{1/2} \rightarrow 4d_{3/2}$  & 9.046 &  8.07(2)  \\
$5p_{1/2} \rightarrow 5d_{3/2}$  & 0.244 &  1.184(3)  \\
$5p_{1/2} \rightarrow 6d_{3/2}$  & 0.512 &  1.002(3)  \\
$5p_{1/2} \rightarrow 7d_{3/2}$  & 0.447 &  0.75(2)  \\
$5p_{1/2} \rightarrow 8d_{3/2}$  & 0.366 &  0.58(2)  \\
$5p_{1/2} \rightarrow 9d_{3/2}$  & 0.304 &  0.45(1)  \\
$5p_{3/2} \rightarrow 6s_{1/2}$  & 6.186 &  6.048(5)  \\
$5p_{3/2} \rightarrow 7s_{1/2}$  & 1.392 &  1.363(4) \\
$5p_{3/2} \rightarrow 8s_{1/2}$  & 0.726 &  0.714(3) \\
$5p_{3/2} \rightarrow 9s_{1/2}$  & 0.476 &  0.468(2) \\
$5p_{3/2} \rightarrow 10s_{1/2}$ & 0.338 &  0.330(2) \\
$5p_{3/2} \rightarrow 4d_{3/2}$  & 4.082 &  3.65(2)  \\
$5p_{3/2} \rightarrow 5d_{3/2}$  & 0.157 &  0.59(2)  \\
$5p_{3/2} \rightarrow 6d_{3/2}$  & 0.255 &  0.48(2)  \\
$5p_{3/2} \rightarrow 7d_{3/2}$  & 0.217 &  0.355(4) \\
$5p_{3/2} \rightarrow 8d_{3/2}$  & 0.176 &  0.272(3) \\
$5p_{3/2} \rightarrow 9d_{3/2}$  & 0.145 &  0.212(2) \\
$5p_{3/2} \rightarrow 4d_{5/2}$  & 12.24 &  10.96(4) \\
$5p_{3/2} \rightarrow 5d_{5/2}$  & 0.493 &  1.76(3)  \\
$5p_{3/2} \rightarrow 6d_{5/2}$  & 0.778 &  1.42(3)  \\
$5p_{3/2} \rightarrow 7d_{5/2}$  & 0.658 &  1.06(2)  \\
$5p_{3/2} \rightarrow 8d_{5/2}$  & 0.530 &  0.81(1)  \\
$5p_{3/2} \rightarrow 9d_{5/2}$  & 0.417 &  0.593(5) \\
\end{tabular}   
\end{ruledtabular}
\end{table}

\subsection{Matrix elements}\label{sec4i}

The matrix elements of Rb atom have been reported several times previously \cite{sahoo1,life7,QC2,arora1,schmieder1,marinescu,zhu,arora3}.
We present these results from our calculations in Table \ref{e1mat} using the
DF and CCSD(T) methods; the differences in the results imply the amount of correlation effects involved to evaluate these matrix elements. We also give uncertainties in the CCSD(T) results mentioned
in the parentheses in the same table. The contributions to these uncertainties
come from the neglected triple excitations in the RCC method and from the incompleteness
of the used basis functions. The uncertainty contribution from the former is estimated from the
differences between the CCSD and CCSD(T) results. Some of the important matrix elements 
are determined more precisely below from the available experimental lifetime results 
of atomic states involving only one (strong) transition channel. However in case
there are more than one (strong) decay channels associated with an atomic state, it
would be intricate to obtain the matrix elements precisely but have been done by optimizing these values 
to reproduce the experimental lifetimes in conjunction with the experimental static
polarizabilities of different atomic states as discussed below.

\begin{table}
\caption{\label{p5s0} Scalar polarizability of the $5s$ state in Rb (in au).
Uncertainties in the results are given in the parentheses.}
\begin{ruledtabular}
\begin{tabular}{lcc}
Contribution & E1 amplitude & Contribution to $\alpha_v^0$ \\
\hline
$\alpha_{5s_{1/2}}(v)$     \\
$5s_{1/2} \rightarrow 5p_{1/2}$  & 4.227(6) &  103.92(1)  \\ 
$5s_{1/2} \rightarrow 6p_{1/2}$  & 0.342(2) &  0.361  \\
$5s_{1/2} \rightarrow 7p_{1/2}$  & 0.118(1) &  0.037  \\
$5s_{1/2} \rightarrow 8p_{1/2}$  & 0.061(5) &  0.009  \\
$5s_{1/2} \rightarrow 9p_{1/2}$  & 0.046(3) &  0.005   \\
\\
$5s_{1/2} \rightarrow 5p_{3/2}$  & 5.977(9) &  203.92(4)   \\
$5s_{1/2} \rightarrow 6p_{3/2}$  & 0.553(3) &  0.940  \\
$5s_{1/2} \rightarrow 7p_{3/2}$  & 0.207(2) &  0.112 \\
$5s_{1/2} \rightarrow 8p_{3/2}$  & 0.114(2) &  0.032  \\
$5s_{1/2} \rightarrow 9p_{3/2}$  & 0.074(2) &  0.013   \\
\\
$\alpha_{5s_{1/2}}(c)$     &       &      9.1(5)      \\
$\alpha_{5s_{1/2}}(vc)$     &       &     $ -0.26(2)$      \\
$\alpha_{5s_{1/2}}(\rm{tail})$     &       &     0.11(1)      \\
\\[0.5pc]
Total                    &       &  318.3(6)    \\    
\end{tabular}   
\end{ruledtabular}
\end{table}

\begin{table}
\caption{\label{p5s1} Dynamic polarizability of the $5s$ state in Rb (in au) at $\lambda = 1064 \ nm$. Uncertainties in the results are given in the parentheses.}
\begin{ruledtabular}
\begin{tabular}{lcc}
Contribution & E1 amplitude & Contribution to $\alpha_v^0$ \\
\hline
$\alpha_{5s_{1/2}}(v)$     \\
$5s_{1/2} \rightarrow 5p_{1/2}$  & 4.227(6) &  235.24(3)  \\
$5s_{1/2} \rightarrow 6p_{1/2}$  & 0.342(2) &  0.428  \\
$5s_{1/2} \rightarrow 7p_{1/2}$  & 0.118(1) &  0.041  \\
$5s_{1/2} \rightarrow 8p_{1/2}$  & 0.061(5) &  0.010  \\
$5s_{1/2} \rightarrow 9p_{1/2}$  & 0.046(3) &  0.006   \\
\\[0.5pc]
$5s_{1/2} \rightarrow 5p_{3/2}$  & 5.977(9) &  441.14(8)   \\
$5s_{1/2} \rightarrow 6p_{3/2}$  & 0.553(3) &  1.114  \\
$5s_{1/2} \rightarrow 7p_{3/2}$  & 0.207(2) &  0.127 \\
$5s_{1/2} \rightarrow 8p_{3/2}$  & 0.114(2) &  0.035  \\
$5s_{1/2} \rightarrow 9p_{3/2}$  & 0.074(2) &  0.014   \\
\\[0.5pc]
$\alpha_{5s_{1/2}}(c)$     &       &      9.3(5)      \\
$\alpha_{5s_{1/2}}(vc)$     &       &     -0.26(2)      \\
$\alpha_{5s_{1/2}}(\rm{tail})$    &       &  0.12(1)     \\
\\[0.5pc]
Total                    &       & 687.3(5)     \\
Experiment \cite{bonin}               &       &  769(61) \\
\end{tabular}
\end{ruledtabular}
\end{table}

In order to evaluate the magnitude of the $5s \rightarrow 5p_{1/2}$ E1 transition 
matrix element, we use the measured lifetime of the $5p_{1/2}$ state which was 
reported as 27.75(8) $ns$ in Ref. \cite{life8}. Using the fact that the $5p_{1/2}$ 
state decays only to the $5s$ state, the line strength of the
$5s \rightarrow 5p_{1/2}$ transition can be obtained by combining this measured lifetime 
with the experimental wavelength ($\lambda=7949.8$ \AA) of the
corresponding transition. The value of the E1 matrix element of the 
$5s \rightarrow 5p_{1/2}$ transition is obtained from this result as 4.227(6) au.
Similarly, it is possible to deduce the magnitude of the E1 matrix element of the
$5s \rightarrow 5p_{3/2}$ transition by combining the measured lifetime of the $5p_{3/2}$
state, reported as 26.25(8) $ns$ \cite{life8}, with its experimental wavelength
(7802.4 \AA). However, the $5p_{3/2}$ state has non-zero transition probabilities to
the $5s$ and $5p_{1/2}$ states via the allowed E1 and the forbidden M1 and E2 
 channels. We found from our calculations that the transition
probabilities through the forbidden channels are very small and negligibly influence the 
lifetime of the $5p_{3/2}$ state; in fact lies within the reported experimental
error bar. Neglecting these contributions, we extract the E1 matrix element of the
$5s \rightarrow 5p_{3/2}$ transition to be 5.977(9) au. 

\begin{table}
\caption{\label{p5p10} Scalar polarizability of the $5p_{1/2}$ state in Rb (in au). Uncertainties in the results are given in the parentheses.}
\begin{ruledtabular}
\begin{tabular}{lcc}
Contribution & E1 amplitude & Contribution to $\alpha_v^0$ \\
\hline
$\alpha_{5p_{1/2}}(v)$     \\
$5s_{1/2} \rightarrow 5p_{1/2}$  & 4.227(6) &  $-103.92(1)$  \\ 
$5p_{1/2} \rightarrow 6s_{1/2}$  & 4.144(3) &  166.32(1) \\
$5p_{1/2} \rightarrow 7s_{1/2}$  & 0.962(4) &  4.93 \\
$5p_{1/2} \rightarrow 8s_{1/2}$  & 0.507(3) &  1.14 \\
$5p_{1/2} \rightarrow 9s_{1/2}$  & 0.333(1) &  0.452 \\
$5p_{1/2} \rightarrow 10s_{1/2}$ & 0.235(1) &  0.215 \\
\\
$5p_{1/2} \rightarrow 4d_{3/2}$  & 8.069(2)  & 702.89(3) \\
$5p_{1/2} \rightarrow 5d_{3/2}$  & 1.184(3) &  7.816(1)  \\
$5p_{1/2} \rightarrow 6d_{3/2}$  & 1.002(3) &  4.560  \\
$5p_{1/2} \rightarrow 7d_{3/2}$  & 0.75(2) &   2.325(1) \\
$5p_{1/2} \rightarrow 8d_{3/2}$  & 0.58(2) &   1.320(1) \\
$5p_{1/2} \rightarrow 9d_{3/2}$  & 0.45(1) &   0.770 \\
$\alpha_{5p_{1/2}}(c)$     &       &      9.1(5)      \\
$\alpha_{5p_{1/2}}(vc)$     &       &      $\sim0.0$      \\
$\alpha_{5p_{1/2}}(\rm{tail})$     &       &      12.6(1.0)      \\
\\[0.5pc]
Total                    &       &   810.5(1.1) \\
\end{tabular}   
\end{ruledtabular}
\end{table}

The estimated E1 matrix elements for the $5s-5p$ transitions from the experimental data are in close agreement
with our calculated results within the predicted uncertainties. These results are further used, along with other matrix elements obtained from the CCSD(T) method, 
to calculate of polarizabilities of the $5s$ and $5p$ states. In Table \ref{p5s0}, we list the the polarizability of the $5s$ state as
318.3(6) au along with the detailed breakdown of the various contributions. The most precise experimental result
reported for this quantity as 318.79(1.42) au \cite{spol} is in excellent agreement with our result. As shown in Table \ref{p5s0}, the dominant
contributions to the $5s$ state polarizability are from the $5s-5p$ transitions following a significant
contribution from the core correlation. We have calculated core correlation contribution
using the MBPT(3) method and the given uncertainty is estimated by scaling the 
wave functions. Our result for the core contribution is  in very good agreement
with the result obtained using the random phase approximation (RPA) \cite{datatab2}. 
Consistency in the estimated $5s$ polarizability value obtained using the 
$5s \rightarrow 5p_{1/2}$ and $5s \rightarrow 5p_{3/2}$ matrix elements, which are obtained 
from the experimental lifetimes of the $5p$ states, and 
experimental polarizability result suggests that both these matrix elements are very accurate.
In order to test the accuracy of our results further we reproduce the
dynamic polarizability of the $5s$ state at $\lambda=1064 \ nm$ whose experimental
value is reported as 769(61) au \cite{bonin}. As shown in Table \ref{p5s1} our result shows a large discrepancy with the experimental measurement. 
Even after replacing the above E1 matrix elements with the calculated CCSD(T) results,
which are slightly larger in magnitude, the polarizability result still does not agree within the experimental
error bar. Therefore, it would be instructive to perform another measurement of this dynamic
polarizability to assert this result.

\begin{table}
\caption{\label{p5p30} Scalar and tensor polarizabilities of the $5p_{3/2}$ state in Rb (in au). Uncertainties in the results are given in the parentheses.}
\begin{ruledtabular}
\begin{tabular}{lccc}
Contribution & E1 amplitude & $\alpha_v^0$ & $\alpha_v^2$\\
\hline
$\alpha_{5p_{1/2}}(v)$     \\
$5p_{3/2} \rightarrow 5s_{1/2}$  & 5.977(9) &  $-101.96(2)$  & 101.96(2) \\ 
$5p_{3/2} \rightarrow 6s_{1/2}$  & 6.048(5) &  182.89(2) & $-182.89(2)$ \\
$5p_{3/2} \rightarrow 7s_{1/2}$  & 1.363(4) &  5.036 & $-5.036$ \\
$5p_{3/2} \rightarrow 8s_{1/2}$  & 0.714(3) &  1.149 & $-1.149$ \\
$5p_{3/2} \rightarrow 9s_{1/2}$  & 0.468(2) &  0.453 & $-0.453$ \\
$5p_{3/2} \rightarrow 10s_{1/2}$ & 0.330(2) &  0.215 & $-0.215$ \\ \\
$5p_{3/2} \rightarrow 4d_{3/2}$  & 3.65(2) & 74.52(1) & 59.62(1) \\
$5p_{3/2} \rightarrow 5d_{3/2}$  & 0.59(2) &  0.988 & 0.791 \\
$5p_{3/2} \rightarrow 6d_{3/2}$  & 0.48(2) &  0.531 & 0.425 \\
$5p_{3/2} \rightarrow 7d_{3/2}$  & 0.355(4) &  0.264 & 0.211 \\
$5p_{3/2} \rightarrow 8d_{3/2}$  & 0.272(3) &  0.147 & 0.118 \\
$5p_{3/2} \rightarrow 9d_{3/2}$  & 0.212(2) &  0.086 & 0.069 \\
\\
$5p_{3/2} \rightarrow 4d_{5/2}$  & 10.89(1) & 663.4(5) & $-132.7(1)$ \\
$5p_{3/2} \rightarrow 5d_{5/2}$  & 1.76(3) &  8.792(5) & $-1.758(1)$  \\
$5p_{3/2} \rightarrow 6d_{5/2}$  & 1.42(3) &  4.647(3) & $-0.929(1)$  \\
$5p_{3/2} \rightarrow 7d_{5/2}$  & 1.06(2) &  2.353(1) & $-0.471$ \\
$5p_{3/2} \rightarrow 8d_{5/2}$  & 0.81(1) &  1.304 & $-0.261$ \\
$5p_{3/2} \rightarrow 9d_{5/2}$  & 0.593(5) & 0.677 & $-0.135$ \\
$\alpha_{5p_{3/2}}(c)$     &       &      9.1(5)     & 0.0  \\
$\alpha_{5p_{3/2}}(vc)$     &       &      $\sim0.0$  & $\sim0.0$    \\
$\alpha_{5p_{3/2}}(\rm{tail})$     &       &     13.40(1.5)  & $-3.15(50)$    \\
\\[0.5pc]
Total                    &       &   868.0(1.7) & $-165.9(5)$ \\
\end{tabular}   
\end{ruledtabular}
\end{table}

It seems from the above analysis that the  calculated E1 matrix 
elements using the CCSD(T) method are reasonably accurate and can be
further employed to obtain the polarizabilities of the $5p$ states. 
However, we can calculate the polarizabilities of the $5p$ states even 
more precisely if the uncertainties in the dominant contributing E1 matrix
elements of the $6s \rightarrow 5p_{1/2,3/2}$, $4d_{3/2} \rightarrow 5p_{1/2,3/2}$ and 
$4d_{5/2} \rightarrow 5p_{3/2}$ transitions are pushed down further.
In order to do so, we evaluate the lifetime of the $6s$ state using our calculated
matrix elements as 4.144(3) and 6.048(5) in au of the $6s-5p_{1/2}$ and 
$6s-5p_{3/2}$ transitions, respectively. We obtain its lifetime as 45.44(8) $ns$
against the experimental result 45.57(17) $ns$ \cite{life3} with branching ratios 
34\% to the $5p_{1/2}$ state and 66\% to the $5p_{3/2}$ state neglecting the observed 
insignificant transition probabilities to the $5s$ and $4d$ states. Since we are able
to obtain more precise lifetime for the $6s$ state using our calculated matrix elements than the measurement, we assume 
these calculated E1 matrix elements are more precise than what we would have obtained from the known
experimental lifetime result.

We would further like to use the above E1 matrix elements to produce the
experimental polarizability of the $5p_{1/2}$ state from which we anticipate to estimate the
E1 matrix element of the $5p_{1/2} \rightarrow 4d_{3/2}$ transition accurately. Since
no direct measurement of the polarizability of the $5p_{1/2}$ state is known to us from the literature,
we use the differential polarizability of the $5s \rightarrow 5p_{1/2}$ transition which is 
reported as 492.20(7) au \cite{hunter1}. In fact, using the differential polarizability
here is advantageous for the following three reasons: (i) we have already
determined the polarizability of the $5s$ state precisely, (ii) the differential
polarizability is not affected by the uncertainty of the core correlation 
contribution, and (iii) precise values of few important  matrix elements contributing towards 
the $5p$ state polarizability are known to some extent from the above analysis.  By adding the
experimental differential polarizability with the precisely known polarizability of the
$5s$ state we consider the experimental $5p_{1/2}$ state polarizability as 810.6(6) au;
indeed this result will not meddle the above advantages. We find from our calculations 
that the E1 matrix elements of the $5p_{1/2}-5s$, $5p_{1/2}-6s$, and $5p_{1/2}-4d_{3/2}$ transitions
have crucial contributions to the $5p_{1/2}$ state polarizability. Substituting the
precisely known values of the first two elements to reproduce the experimental $5p_{1/2}$ polarizability result, the E1 matrix element of the 
$5p_{1/2} \rightarrow 4d_{3/2}$ transition is set as 8.069(2) au. This agrees well with our calculated result 8.07(2) au.
From this analysis, we estimate theoretical value of the $5p_{1/2}$ state polarizability to be
810.5(1.1) au; contributions from various parts are given explicitly in Table \ref{p5p10}.

\begin{table}
\caption{\label{lifetime}Comparison of lifetimes (in $ns$) of three excited states in Rb atom from various theoretical and experimental studies.} 
\begin{ruledtabular}
\begin{tabular}{llll}
\multicolumn{1}{l}{Level}&
\multicolumn{1}{l}{Reco} &
\multicolumn{1}{l}{Other theory$^a$} &
\multicolumn{1}{l}{Expt.}\\
\hline
$4d_{3/2}$ & 82.30(17)  & 83.0(8) & 86(6)$^b$  \\
$4d_{5/2}$ & 89.32(16)  & 89.4(9) & 94(6)$^b$  \\
$6s_{1/2}$ & 45.44(8)  & 45.4(1) & 45.57(17)$^c$ \\
\end{tabular} 
\end{ruledtabular}
Refs: $^a$~\cite{life7}, $^b$Ref.~\cite{life1},$^c$Ref.~\cite{life3}.
\end{table}

It can be noticed that the $4d_{5/2}$ state has only one allowed decay channel to 
the $5p_{3/2}$ state. Therefore if the lifetime of the $4d_{5/2}$ state is known 
precisely then the E1 matrix element of the $5p_{3/2} \rightarrow 4d_{5/2}$ 
transition can be estimated accurately from this data. There are two experimental
results for the lifetime of the $4d_{5/2}$ state reported as 89.5 \cite{life9} and 
94(6) $ns$ \cite{life1}. From the former result which is the latest, we deduce 
the E1 matrix element of the above transition to be about 10.89 au which reasonably 
agrees with our CCSD(T) result 10.94(6) au. Matrix element obtain from the later lifetime data
gives much lower absolute value with very large uncertainty compared to our calculated result 
which is not of our interest. Similarly, the lifetimes 
of the $4d_{3/2}$ state are reported as 83.4 $ns$ \cite{life9} and 86(6) $ns$ \cite{life1}.
The $4d_{3/2}$ state has two strong allowed decay channels to the $5p_{1/2}$ and $5p_{3/2}$ 
states. By combining the above E1 matrix element for the $5p_{1/2} \rightarrow 4d_{3/2}$ 
transition and the lifetime of the $4d_{3/2}$ state as 83.4 $ns$ (the reason for not
considering the other value is same as cited above), we predict the E1 matrix element
of the $5p_{3/2} \rightarrow 4d_{3/2}$ transition to be about 3.5 au. In order to find this matrix element more precisely, we use the experimental results 
for the scalar and tensor polarizabilities of the $5p_{3/2}$ state which are
reported as 857(10) and $-163(3)$ in au \cite{krenn}, respectively. Our calculation
shows that the major contributions to these polarizabilities come from the 
matrix elements of the $5p_{3/2} \rightarrow 5s_{1/2}$,
$5p_{3/2} \rightarrow 6s_{1/2}$, $5p_{3/2} \rightarrow 4d_{3/2}$
and $5p_{3/2} \rightarrow 4d_{5/2}$ transitions.
From the sensitivity in the given precision of the E1 matrix element
of the $5p_{3/2} \rightarrow 4d_{3/2}$ transition to be able to reproduce the scalar
and tensor polarizabilities in their respective error bars, we get a lower 
bound for this matrix element as 3.6 au. Without any loss of quality, we retain our
CCSD(T) result, i.e. 3.65(2) au, as the most precise value for this matrix element.
Using these optimized results and the combined experimental values of the scalar and 
tensor polarizabilities of the $5p_{3/2}$ state, we get the best value for the E1 matrix 
element of the $5p_{3/2} \rightarrow 4d_{5/2}$ transition to be 10.89(1) au.

Now we list below the optimized E1 matrix elements (in au) obtained from the above analysis apart
from our calculated results as:
\begin{eqnarray}
\langle 5s ||D|| 5p_{1/2} \rangle &=& 4.227(6) \nonumber \\
\langle 5s ||D|| 5p_{3/2}  \rangle &=& 5.977(9) \nonumber \\
\langle 5p_{1/2} ||D|| 4d_{3/2}  \rangle &=& 8.069(2) \nonumber \\
\langle 5p_{3/2} ||D|| 4d_{5/2} \rangle &=& 10.89(1) .
\label{bmatv}
\end{eqnarray}

\subsection{Lifetimes of few excited states}
Since we are now able to estimate some of the E1 matrix elements more precisely
than the previously known results, we would like
to use them further to estimate lifetimes of first few excited states in Rb atom
accurately. The matrix elements for the $\langle 5s ||D|| 5p_{1/2} \rangle$ 
and $\langle 5s ||D|| 5p_{3/2}  \rangle$ transitions were obtained from the lifetime
measurements, so we still consider the most accurately known lifetimes of
the $5p_{1/2}$ and $5p_{3/2}$ states from the experiment as 27.75(8) $ns$ and 26.25(8) $ns$,
respectively. We now determine the lifetimes of the $4d$ and $6s$ states 
using the E1 matrix elements listed in Eq. (\ref{bmatv}) and from our 
calculations which are given in Table \ref{e1mat}. The estimated lifetimes are mentioned
in Table \ref{lifetime} as recommend (Reco) values and compared
with the other available experimental and theoretical results in the 
same table.
\begin{figure}[t]
  \includegraphics[scale=0.3, angle=270]{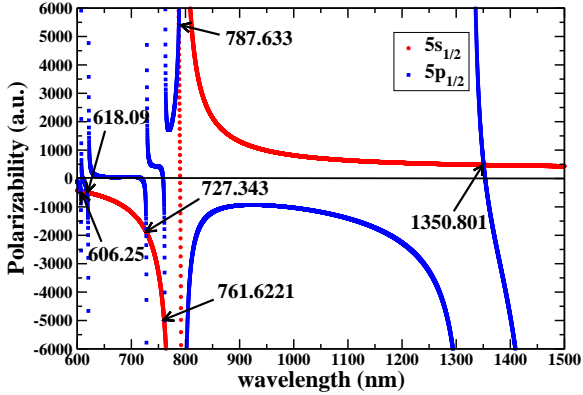}
  \caption{(color online) Magic wavelengths identified by arrows for the $5p_{1/2}-5s$ transition 
in Rb using the linearly polarized light.}
  \label{figrb-1}     
\end{figure} 

\begin{figure}[t]
  \includegraphics[scale=0.3]{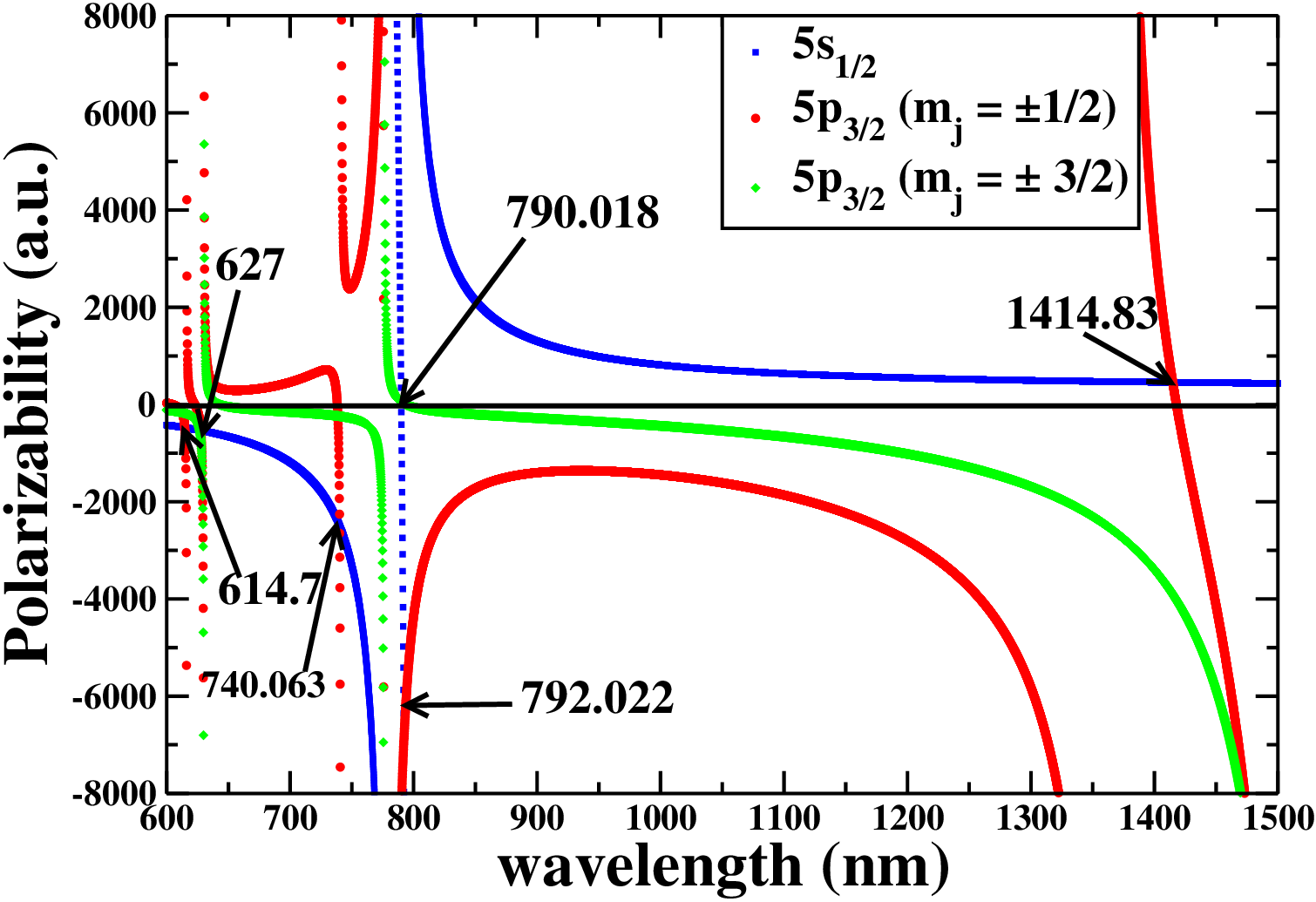}
  \caption{(color online) Magic wavelengths identified by arrows for the $5p_{3/2}-5s$ transition 
in Rb using the linearly polarized light.}
  \label{figrb-2}     
\end{figure}

\begin{table}
\caption{\label{pol-comp} Comparison of the static and ac polarizabilities (in au) in Rb atom 
for the $5s$, $5p_{1/2}$ and $5p_{3/2}$ states with other experiments and
theory.}
\begin{ruledtabular}
\begin{tabular}{llllll}
\multicolumn{1}{l}{ }&
\multicolumn{1}{l}{ $\alpha^0_{5s}$ }&
\multicolumn{1}{l}{$\alpha^0_{5p_{1/2}}$ } &
\multicolumn{1}{l}{$\alpha^0_{5p_{3/2}}$} &
\multicolumn{1}{l}{$\alpha^2_{5p_{3/2}}$}\\
\hline		
Present & 318.3(6) & 810.5(1.1) & 868.0(1.7) & $-165.9(5)$ \\
Other & 317.39$^a$ & 805$^d$ & 867$^d$	& $-167^d$   \\
Other & 318.6(6)$^b$ & 807$^e$ & 870$^e$   & $-171^e$    \\
Exp.	& 318.79(1.42)$^c$ & 810.6(6)$^f$ & 857(10)$^g$ & $-163(3)^g$\\	
\end{tabular} 
\end{ruledtabular}
\begin{tabular}{ll}
References: &  $^a$ \cite{spol2} \\
  & $^b$ \cite{derevianko2} \\
  & $^c$ \cite{spol} \\
  & $^d$ \cite{arora1} \\
  & $^e$ \cite{poldere} \\
  & $^f$ \cite{hunter1,pol-andrei}\\
  & $^g$ \cite{krenn}\\
\end{tabular} 
\end{table}

\subsection{Status of the polarizability results}\label{sec4ii}

To affirm the broad interest of studying polarizabilities in Rb atom, we discuss briefly 
below about various experimental and theoretical works in the evaluation of polarizabilities
of the $5s$ and $5p$ states reported so far in the Rb atom. There were several measurements
carried out on Stark shifts in Rb atom almost two decades ago \cite{hunter1,hunter2,hunter3,
tanner1} from which the polarizabilities of the $5s$ ground state and few excited states
were estimated.
Hunter and coworkers had observed the dc Stark shifts of the D1 line in Rb using a pair of
cavity stabilized diode lasers locked to resonance signals \cite{hunter1,hunter2,hunter3}.
In another work, Tanner and Wieman had used a crossed-beam laser spectroscopy with frequency 
stabilized laser diodes to measure the differential Stark shift of D2 line \cite{tanner1}.
Marrus \textit{et al.} had used a atomic beam method long ago to measure the Stark 
shift from which both the scalar and tensor polarizabilities of the $5p_{3/2}$ state
were determined \cite{marrus}. 

\begin{table}
\caption{\label{tabRb-linear}
Magic wavelengths $\lambda_{\rm{magic}}$ for the linearly polarized light above 600~$nm$ for the 
$5p_{1/2}-5s$ and $5p_{3/2}-5s$ transitions in Rb and the corresponding values of polarizabilities
at the magic wavelengths. The wavelengths (in vacuum) are given in $nm$ and polarizabilities are 
given in au. The given $m_j$ values correspond to the $5p$ states.}
\begin{ruledtabular}
\begin{tabular}{lccccc}
\multicolumn{3}{l}{Transition: $5p_{1/2}~-~5s$} &\\ [0.2pc]
\hline
\multicolumn{1}{c}{}&
\multicolumn{1}{c}{$|m_j|$} &
\multicolumn{1}{c}{$\lambda_{\rm{magic}}$} &
\multicolumn{1}{c}{$\lambda_{\rm{magic}}$Ref.~\cite{arora1}}&
\multicolumn{1}{c}{$\alpha(\lambda_{\rm{magic}})$} \\
\hline
&1/2 & 606.25(1) &606.2(1)  & $-443.3$ \\ 
&1/2 & 618.09(2) & 617.7(7)  & $-490$\\
&1/2 & 727.343(2) & 727.35(1) & $-1876$\\
&1/2 & 761.6221(2) & 761.5(1) & $-5270$ \\
&1/2 & 787.633(2) & 787.6(1)  & 5382 \\
&1/2 & 1350.801(9) & 1350.9(5)&475.5 \\
\hline
\multicolumn{3}{l}{Transition: $5p_{3/2}~-~5s$} \\ [0.2pc]
\hline

\multicolumn{1}{c}{}&
\multicolumn{1}{c}{$|m_j|$} &
\multicolumn{1}{c}{$\lambda_{\rm{magic}}$} &
\multicolumn{1}{c}{$\lambda_{\rm{magic}}$Ref.~\cite{arora1}}&
\multicolumn{1}{c}{$\alpha(\lambda_{\rm{magic}})$} \\
\hline 
& 1/2 &614.70(1) & 614.7(1) &$-477$ \\
\\
& 3/2 & 626.62(3) & 626.2(9) &$-529$ \\
& 1/2 & 627.70(1)& 627.3(5) &$-534$ \\
\\
& 1/2 & 740.063(2) &740.07(1)	&$-2493$ \\
\\
& 1/2 & 775.868(1) &775.84(1) &$-20030$ \\
& 3/2 & 775.8228(2) & 775.77(3) &$-19917$ \\
\\
& 3/2 & 790.018(2) & 789.98(2) & 53 \\
& 1/2  &792.022(1) & 792.00(1)	&$-6973$ \\
\\
& 1/2 & 1414.83(3) &1414.8(5)& 455  \\
\end{tabular}   
\end{ruledtabular}
\end{table}

The extensive calculation of polarizabilities in Rb atom was first carried out by
Marinescu \textit{et al.} using an $l$-dependent model potential \cite{marinescu}.
In this work, the infinite second-order sums in the polarizability calculations
were transformed into integrals over the solutions of two coupled inhomogeneous
differential equations and the integrals were carried out using Numerov integration 
method \cite{numerov}. In 2004, Zhu et al. employed the RCC method to calculate
the scalar and tensor polarizabilities of the ground and the first $p$ excited
states in alkali atoms \cite{zhu}. The results obtained using the RCC method
were substantially improved over the earlier calculations based on the 
non-relativistic theories. Later, Arora \textit{et al.} extended these 
calculations to obtain frequency-dependent scalar and tensor 
polarizabilities of the ground and first excited $5p$ states in Rb 
\cite{arora1,arora2,arora3} using the RCC method at the linearized singles, 
doubles and partial triples excitations level (SDpT method).

As has been discussed earlier, we have optimize at least seven important
E1 matrix elements which are crucial in obtaining the polarizabilities
of the $5s$ and $5p$ states and the other matrix elements have been obtained using
the CCSD(T) method which includes all the non-linear terms. Therefore, the 
predicted polarizabilities of the $5s$ and $5p$ states obtained in this work are expected
to be accurate enough to employ them further in the determination of the 
magic wavelengths in Rb atom, which is the prime motivation of the present 
work. In Table \ref{pol-comp}, we compare our polarizability results with the other reported
values. Our results are more polished over the earlier studied 
results mainly due to the optimization of the matrix elements.

In our knowledge, there are no experimental and/or theoretical results on
vector polarizabilities of the $5s$ and $5p$ states for any wavelengths 
available in Rb atom to compare with the present calculations. Accuracy in
these polarizabilities will determine
correct values of the magic wavelengths for the circularly polarized light
in this atom. Since the E1 matrix elements required to determine the vector
polarizabilities are same as required for the calculation of scalar and tensor
polarizabilities, we expect a similar precision in our used vector polarizability 
results as discussed in the last subsection. We shall present the vector polarizabilities 
in the $5s$ and $5p$ states at a given wavelength (say close to a particular $\lambda_{\rm{magic}}$
value) so that if necessary our results can also be further verified by any other study.

\begin{table}
\caption{\label{Rb0}Contributions to the $5s$ scalar ($\alpha^{0}_v$) and vector 
($\alpha^{1}_v$) polarizabilities at $\lambda$=770~$nm$ in Rb.
Uncertainties in the results are given in the parentheses.}
\begin{ruledtabular}
\begin{tabular}{lrrr}
\multicolumn{1}{c}{Contribution} &
\multicolumn{1}{c}{$\alpha^0_v$}&
\multicolumn{1}{c}{$\alpha^1_v$}\\
\hline
$\alpha_{5s_{1/2}}(v)$ \\
$5s_{1/2} \rightarrow 5p_{1/2}$& $-1576.1(2) $& $3254.4(4)$    \\
$5s_{1/2} \rightarrow 6p_{1/2}$&0.515 &  $-0.565 $  \\
$5s_{1/2} \rightarrow 7p_{1/2}$&0.047  &0.044 \\
$5s_{1/2} \rightarrow 8p_{1/2}$&0.011 & 0.010 \\
$5s_{1/2} \rightarrow 9p_{1/2}$&0.006 & 0.005 \\
\\[0.5pc]
$5s_{1/2} \rightarrow 5p_{3/2}$& $-7615(1) $ &  $-7716(1) $    \\
$5s_{1/2} \rightarrow 6p_{3/2}$&1.339 & 0.731  \\
$5s_{1/2} \rightarrow 7p_{3/2}$&0.144 & 0.067\\
$5s_{1/2} \rightarrow 8p_{3/2}$&0.039 & 0.017 \\
$5s_{1/2} \rightarrow 9p_{3/2}$&0.016 & 0.007 \\
[0.5pc]\\
$\alpha_{5s_{1/2}}(c)$ &  9.2(5) & 0.0 & \\
$\alpha_{5s_{1/2}}(vc)$ & $-0.26(2)$ & $\sim 0.0$  \\
$\alpha_{\rm{tail}}$&  0.14(1)   &  0.002(1)      \\
Total               &  $-9180(1.1) $ & $-4462(1.1)$        
\end{tabular}   
\end{ruledtabular}
\end{table}

\subsection{AC Stark shifts and magic wavelengths}\label{sec4iii}

Following Eq. (\ref{cpl}), the ac Stark shift $\Delta E_v$ of an atomic energy level
$E_v$ due to the external applied ac electric field $\mathcal{E}$, in the absence of 
any magnetic field, can be parametrized in terms of $\alpha_0$, $\alpha_1$ and $\alpha_2$ as \cite{derevianko1}
\begin{eqnarray}
	\Delta E_v &=& -\frac{1}{2}\mathcal{E}^2 [ \alpha_v^0(\omega)
	+\mathcal{A} \frac{m_j}{j_v} \alpha_v^1(\omega) \nonumber \\ &&
	+ \left(\frac{3m_j^2-j_v(j_v+1)}{j_v(2j_v-1)} \right)\alpha_v^2(\omega) ]
\label{eq-pol}
\end{eqnarray}
In this formula, the frequency $\omega$ is assumed to be several line-widths off-resonance.
The differential ac Stark shift for a transition is defined as the difference between the Stark shifts
of individual levels. For instance, the interested differential ac Stark shifts in our case
are for the $5p_i-5s$ transitions (with $i=1/2,3/2$) which are given by
\begin{eqnarray}
 \delta(\Delta E)_{5p_i-5s} &=&  \Delta E_{5p_i} - \Delta E_{5s} \nonumber \\
   &=& \frac{1}{2} \mathcal{E}^2 (\alpha_{5s} - \alpha_{5p_i}),
\end{eqnarray}
where we have used the total polarizabilities of the respective states. Since
the external electric field $\mathcal{E}$ is arbitrary, we can verify the frequencies or wavelengths where
$\alpha_{5p_i}=\alpha_{5s}$, for the null differential ac Stark shifts.

In order to estimate the total polarizability for any particular set of $j_v$ and $m_j$ values, 
we need to determine the scalar, vector and tensor polarizabilities.
Magic wavelengths are calculated for a 
continuous values of frequencies (can also be expressed in terms of
wavelength $\lambda$) by plotting the total polarizability for different states against the
$\lambda$ values. The crossing between the two polarizabilities at various values 
of wavelengths will correspond to $\lambda_{\rm{magic}}$. Trapping of Rb atoms is 
convenient at these wavelengths as was stated in the beginning.
As pointed out in \cite{arora1}, the linearly polarized lattice scheme
offers only a few cases in which the magic wavelengths are suitable
from the experimental point of view. Therefore, we would like to explore
the idea of using the circularly polarized light for which the magic
wavelengths need to be determined separately for each magnetic quantum 
number $m_j$.

\begin{table}
\caption{\label{Rb1}Contributions to the $5p_{1/2}$ scalar ($\alpha^{0}_v$) and vector ($\alpha^{1}_v$) polarizabilities at $\lambda$=770~$nm$ in Rb.
Uncertainties in the results are given in the parentheses. }
\begin{ruledtabular}
\begin{tabular}{lrr}
\multicolumn{1}{c}{Contribution} &
\multicolumn{1}{c}{$\alpha^0_v$}&
\multicolumn{1}{c}{$\alpha^1_v$}\\
\hline
$\alpha_{5p_{1/2}}(v)$ \\
$5p_{1/2} \rightarrow 5p_{1/2}$ &   1567.1(2) & 3254.4(4) \\
$5p_{1/2} \rightarrow 6s_{1/2}$ &  $-85.029(5) $  &292.38(2)  \\
$5p_{1/2} \rightarrow 7s_{1/2}$ & 46.676(4) & $-88.283(7)$ \\
$5p_{1/2} \rightarrow 8s_{1/2}$ & 3.020 & $-4.764$ \\
$5p_{1/2} \rightarrow 9s_{1/2}$ & 0.954 & $-1.382$ \\
$5p_{1/2} \rightarrow 10s_{1/2}$ &0.412 &  $-0.570$
\\[0.5pc]
$5p_{1/2} \rightarrow 4d_{3/2}$ &  $-262.99(1) $&  $-504.00(2)$ \\
$5p_{1/2} \rightarrow 5d_{3/2}$ & 382.94(2)&  379.02(2)  \\
$5p_{1/2} \rightarrow 6d_{3/2}$ & 13.029(1) &  10.504(1)\\
$5p_{1/2} \rightarrow 7d_{3/2}$ & 5.035(2) & 3.694(2) \\
$5p_{1/2} \rightarrow 8d_{3/2}$ & 2.565(1) &  1.787(1) \\
$5p_{1/2} \rightarrow 9d_{3/2}$ & 1.413 &  0.954 \\
[0.5pc]\\
$\alpha_{5p_{1/2}}(c)$ & 9.2(5) & 0.0 \\
$\alpha_{5p_{1/2}}(vc)$ & $\sim 0.0$ & $\sim 0.0$  \\
$\alpha_{\rm{tail}}$&  17.6(20)  &    3.8(4)      \\
Total               &  1711(2) & 3347.7(4)         
\end{tabular}   
\end{ruledtabular}
\end{table}

In the next two subsections, we shall discuss about the magic wavelengths
for the $5p_{1/2,3/2}-5s$ transitions for both the linearly and circularly
polarized lights. The reason for bringing up the issue of magic wavelengths
for the linearly polarized lights in these transitions is that since we have obtained
the most accurate results for all the static polarizabilities it is expected 
that we will get better results for the magic wavelengths using our optimized 
set of E1 matrix elements. This
will also help us in making a comparison study between the results
obtained from the linearly and circularly polarized lights.

\subsection{Case for the linearly polarized optical traps}\label{sec4iv}

Since we are interested in optical traps and the previous study \cite{arora1} 
reveals that the magic wavelengths for the $5s-5p$ transitions at which the Rb
atom can be trapped using the linearly polarized lights lie in between $600-1500 \ nm$,
we try to find out the null differential polarizabilities in this
region. In Fig. \ref{figrb-1}, we plot the total polarizabilities due to the linearly
polarized lights for both the $5s$ and $5p_{1/2}$ states. As seen in the
figure, the $5s$ state dynamic polarizabilities are generally small in  this
region except for the wavelengths in close vicinity to the $5s-5p_{1/2}$ resonance (at 795 $nm$) and $5s-5p_{3/2}$
resonance (at 780 $nm$). However, the $5p_{1/2}$ state has several resonances
in the  considered wavelength range. It is generally expected that the
$5p_{1/2}$ state polarizability will cross the $5s$ state polarizability 
in between each pair of resonances. We found total six magic wavelengths
for the $5p_{1/2} \rightarrow 5s$ transition in between the five resonances.

\begin{table}
\caption{\label{Rb2}Contributions to the $5p_{3/2}$ scalar ($\alpha^{0}_v$) and vector ($\alpha^{1}_v$) and tensor ($\alpha^{2}_v$) polarizabilities at $\lambda$=770~$nm$ in Rb.
Uncertainties in the results are given in the parentheses. }
\begin{ruledtabular}
\begin{tabular}{lrrrrr}
\multicolumn{1}{c}{Contribution} &
\multicolumn{1}{c}{$\alpha^0_v$}&
\multicolumn{1}{c}{$\alpha^1_v$}&
\multicolumn{1}{c}{$\alpha^2_v$}\\
\hline
$\alpha_{5p_{3/2}}(v)$ \\
$5p_{3/2} \rightarrow 5s_{1/2}$ & 3807.5(7) & 11575(2) & $-3807.5(7)$ \\
$5p_{3/2} \rightarrow 6s_{1/2}$ & $-85.017(9)$ & 452.76(5) & 85.017(9)\\
$5p_{3/2} \rightarrow 7s_{1/2}$ &68.181(6) & $-196.85(2)$ &$-68.181(6)$\\ 
$5p_{3/2} \rightarrow 8s_{1/2}$ &3.194 & $-7.667(1)$ &$-3.194$\\
$5p_{3/2} \rightarrow 9s_{1/2}$ &0.984 & $-2.168$ &$-0.984$\\
$5p_{3/2} \rightarrow 10s_{1/2}$ &0.422 &  $-0.885$ & $-0.422$
\\[0.5pc]
$5p_{3/2} \rightarrow 4d_{3/2}$ & $-25.311(5)$&60.32(1)& $-20.249(4)$\\
$5p_{3/2} \rightarrow 5d_{3/2}$ & $-61.57(1)$&74.47(1) & $-49.25(1)$\\
$5p_{3/2} \rightarrow 6d_{3/2}$ & 1.607(1)& $-1.578(1)$ &1.286(1)\\
$5p_{3/2} \rightarrow 7d_{3/2}$ & 0.591&  $-0.527$  &0.473\\
$5p_{3/2} \rightarrow 8d_{3/2}$ & 0.293 &  $-0.248$  & 0.234 \\
$5p_{3/2} \rightarrow 9d_{3/2}$ & 0.162 &  $-0.133$ &  0.130
\\[0.5pc]
$5p_{3/2} \rightarrow 4d_{5/2}$ & $-225.3(2)$ & $-805.4(6)$ & 45.06(4)   \\
$5p_{3/2} \rightarrow 5d_{5/2}$ & $-564.1(3)$ &  $-1023.3(6)$ &112.8(1) \\
$5p_{3/2} \rightarrow 6d_{5/2}$ & 14.06(1)&  20.70(1) &$-2.811(2)$\\
$5p_{3/2} \rightarrow 7d_{5/2}$ & 5.264(2) &  7.046(3)   &$-1.053$  \\
$5p_{3/2} \rightarrow 8d_{5/2}$ & 2.597(1) &  3.298(1)  & $-0.519$\\
$5p_{3/2} \rightarrow 9d_{5/2}$ &  1.2700 &   1.562 & $-0.254$  
\\[0.5pc]
$\alpha_{5p_{3/2}}(c)$ & 9.3(5) & 0.0&0.0 \\
$\alpha_{5p_{3/2}}(vc)$ & $\sim 0.0$ & $\sim 0.0$  & $\sim 0.0$\\
$\alpha_{\rm{tail}}$&  19(2) &    6.9(7)  & $-4.7(9)$ \\
Total               &  2973(2)& 10163(5) &  $-3714(1)$      
\end{tabular}   
\end{ruledtabular}
\end{table}

However, the case for the $5p_{3/2} \rightarrow 5s$ transition is different owing
to the presence of non-zero tensor contribution of the $5p_{3/2}$ state.
As shown in Fig. \ref{figrb-2}, we get different magic wavelengths 
for the $5p_{3/2} \rightarrow 5s$ transition at $m_j=\pm 1/2$ and 
$m_j=\pm 3/2$ sub-levels of the $5p_{3/2}$ state. There are few wavelengths in between resonances
where $\alpha_{5p_{3/2}}$ with $m_j=\pm3/2$ contribution is not
same as the $\alpha_{5s}$. This leads to reduction in the number of magic wavelengths for this transition.
For example, we did not find any $\lambda_{\rm{magic}}$ between the 
$5p_{3/2}-4d_{3/2,5/2}$ resonances (at 1529 $nm$) and the
$5p_{3/2}-6s$ resonance (at 1367 $nm$) for $m_j=\pm 3/2$ sublevels of the $5p_{3/2}$ state.

We have limited our search for the magic wavelengths where the differential
polarizabilities between the $5s$ and $5p_{j}$ states are less than 0.5\%. Based on all these data, we list 
now $\lambda_{\rm{magic}}$ (in vacuum) above 600 $nm$ in Table 
\ref{tabRb-linear} for the $5p_{1/2}-5s$ and $5p_{3/2}-5s$ transitions
in Rb atom and compare them with the previously known results. The
present results are improved slightly due to the optimized E1
matrix elements used here. The uncertainties in our magic wavelength results are found as the maximum differences between the
$\alpha_{5s} \pm \delta\alpha_{5s}$ and $\alpha_{5p} \pm \delta\alpha_{5p}$
contributions with their respective magnetic quantum numbers, where 
the $\delta\alpha$ are the uncertainties in the polarizabilities
for their corresponding states. 

The reason for not acquiring sufficient number of magic wavelengths for the $5p_{3/2}-5s$ 
transition lies in the fact that extra contribution from the  tensor
polarizability to the total $5p_{3/2}$ polarizability is not compensated by the counter
part of the $5s$ state. The idea of using the
circularly polarized light to obtain magic wavelengths
for the $5p_{3/2}-5s$ transition is triggered from that fact that the extra contribution from the tensor polarizability to the 
$5p_{3/2}$ state might be cancelled by the vector polarizability 
contributions or the vector polarizabilities are so large that
they may play a dominant role in determining the differential
polarizabilities. This would be evident in the following subsection.

\begin{figure}[t]
  \includegraphics[scale=0.3, angle=270]{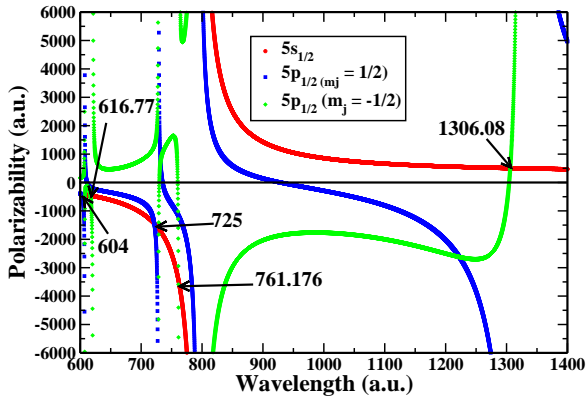}
  \caption{(color online) Magic wavelengths identified by arrows for the $5p_{1/2}-5s$ transition 
in Rb using the left-handed circularly polarized light.}
  \label{figrb-3}     
\end{figure}

\begin{table}
\caption{\label{tabRb-circular1} 
Magic wavelengths $\lambda_{\rm{magic}}$ above 600~$nm$ for the 
 $5p_{1/2} - 5s$  transition in Rb and the 
corresponding values of total polarizabilities at the magic wavelengths for 
the left-circularly polarized laser beam. The wavelengths (in vacuum) 
are given in $nm$ and polarizabilities are given in au. The given $m_j$ 
values are for the $5p$ states.}
\begin{ruledtabular}
\begin{tabular}{lcccc}
\multicolumn{3}{l}{Transition: $5p_{1/2}~-~5s$} &\\ [0.2pc]
\hline
\multicolumn{1}{c}{$m_j$} &
\multicolumn{1}{c}{$\lambda_{\rm{magic}}$} &
\multicolumn{1}{c}{$\alpha(\lambda_{\rm{magic}})$}&
\multicolumn{1}{c}{$\lambda_{\rm{magic}}(\rm{avg})$} \\
\hline
1/2 & 600.83(14) & $-405$ & \\
    &            &  &604(7) \\
-1/2 & 607.98(1) & -428 & \\
\\ \\
-1/2 & 616.77(2) & $-461$ & 617\\
\\\\
1/2 & 721.628(23) & $-1449$ &\\
 & & & 725(7) \\
-1/2 & 728.843(1) & -1633 & \\
\\\\
-1/2 & 761.176(1)& $-3424$ & 761\\
\\ \\
-1/2 & 1306.08(1) & 504 & 1306\\
\end{tabular}   
\end{ruledtabular}
\end{table}

\begin{table}
\caption{\label{tabRb-circular2} 
Magic wavelengths $\lambda_{\rm{magic}}$ above 600~$nm$ for the 
 $5p_{3/2}-5s$ transition in Rb and the 
corresponding values of total polarizabilities at the magic wavelengths for 
the left-circularly polarized laser beam. The wavelengths (in vacuum) 
are given in $nm$ and polarizabilities are given in au. The given $m_j$ 
values are for the $5p$ states.}
\begin{ruledtabular}
\begin{tabular}{lcccc}
\multicolumn{3}{l}{Transition: $5p_{3/2}~-~5s$} &\\ [0.2pc]
\hline
\multicolumn{1}{c}{$m_j$} &
\multicolumn{1}{c}{$\lambda_{\rm{magic}}$} &
\multicolumn{1}{c}{$\alpha(\lambda_{\rm{magic}})$} &
\multicolumn{1}{c}{$\lambda_{\rm{magic}}(\rm{avg})$} \\
\hline
1/2 & 613.25(3) & $-447$& \\
	&	&	&  \\
-1/2 & 615.51(1) & $-456$ & 616(5) \\
	&	&	& \\
-3/2 & 618.15(2) & $-466$& \\
\\ \\
 3/2 & 630.142(1)  & $-516$& \\
	&	&	& \\
1/2 & 628.30(1)  & $-508$ &\\
   &   &  &  628(5)\\
-1/2 & 626.95(1)  & $-502$ &\\
	&	&	& \\
-3/2 & 625.04(3) & $-494$& \\  
\\ \\
 3/2 & 746.737(15) & $-2328$ &\\
	&	&	& \\
 1/2 & 738.794(32) & $-1964$ &\\
   &   &  &  742(8)\\
 -1/2 & 740.587(1) & $-2037$ &\\
	&	&	& \\
 -3/2 & 742.262(1) & $-2109$ &\\
\\ \\
 3/2 & 775.836(5) & $-6231$ &\\
	&	&	& \\
 1/2 & 775.834(7) & $-6230$ &\\
& & & 775.8(2) \\
 -1/2 & 775.789(3) & $-6215$ &\\
	&	&	& \\
 -3/2 &775.693(2) & $-6183$ &\\
\\
 1/2 & 783.883(13) & $-10925$ & \\
	&	&	& \\
 -1/2  & 787.547(4) & $-16431$ & 786(4)\\
	&	&	& \\
-3/2 & 776.497(4) & $-16318$ \\ 
\\
1/2 & 1454.4(9) & 453 &\\
   &   &  &  &\\
-1/2 & 1387.1(1) & 473 &1382(149)\\
	&	&	& \\
-3/2 & 1305.9(1) & 504 &\\
\end{tabular}   
\end{ruledtabular}
\end{table}

\subsection{Case for the circularly polarized optical traps}\label{sec4v}

As mentioned previously, polarizabilities for the circularly polarized light have 
extra contribution from the vector component of the tensor product between the 
dipole operators. This extra factor is expected to provide better results for 
state-insensitive trapping. First, we present the scalar, vector and tensor
dynamic polarizabilities of the $5s$, $5p_{1/2}$ and $5p_{3/2}$ states in Tables
\ref{Rb0}, \ref{Rb1} and \ref{Rb2}, respectively, at $\lambda=770 \ nm$ to perceive
their general behavior. The choice of this wavelength is deliberate for being 
close to one of the magic wavelengths for the circularly polarized light (e.g. see Table
(\ref{tabRb-circular1}) and (\ref{tabRb-circular2})). Hereafter we shall consider the
left-handed circularly polarized light for all the practical purposes as
the results will have a similar trend with the right-handed circularly polarized 
light due to the linear dependency of degree of polarizability $\mathcal{A}$
in Eq. (\ref{eq-pol}). Nevertheless, the left or right handed polarization
in the experimental set up is just a matter of choice.

For the sake of completeness of our study, we also search for magic wavelengths
in the $5s-5p_{1/2}$ transition in Rb atoms using the circularly polarized
light although a fairly large number of magic wavelengths for this transition is
found using the linearly polarized light. For this purpose, we plot net dynamic 
polarizability results of the $5s$ and $5p_{1/2}$ states in Fig. \ref{figrb-3}
using the circularly polarized light against different values of wavelength.
The figure shows that the total polarizability of the $5s$ state for any values 
of $\lambda$ is very small except for the wavelengths 
close to the two primary resonances. Due to the $m_j$ dependence of the vector 
polarizability coefficient in Eq. (\ref{eq-pol}), the crossing occurs at a different 
wavelength for the different values of $m_j$ in between two $5p_{1/2}$ resonances.
As shown in Table \ref{tabRb-circular1}, we get set of five magic wavelengths in between 
seven $5p_{1/2}$ resonances lying in the wavelength range 600-1400 $nm$. Out of
these five sets of magic wavelengths three sets of the magic wavelengths occur only for 
negative values of $m_j$. Thus, the number of convenient magic wavelengths for the above 
transition is less than the number of magic wavelengths obtained for the linearly polarized 
light. This advocates for the use of linearly polarized light in this transition, though
choice of the circularly polarized light is not bad at all. The $m_j$ dependence
of traps and the difficulties in building a viable experimental set up in the case of 
circularly polarized light could be the other major concern.

In this work, we also propose the use of "switching trapping scheme" 
(described below) which may solve the problem in cases where
state-insensitive trapping is only supportive for the negative $m_j$ sublevels 
of $5p$ states. We observed that the
same magic wavelength will support state-insensitive trapping for negative $m_j$ sublevels 
if we switch the sign of $\mathcal{A}$ and $m_j$ of $5s$ state. In other words, the change 
of sign of $\mathcal{A}$ and $m_j$ sublevels of $5s$ state will lead to the same result 
for the positive values of $m_j$ sub-levels of $5p$ states.

\begin{figure}[t]
  \includegraphics[scale=0.3, angle=270 ]{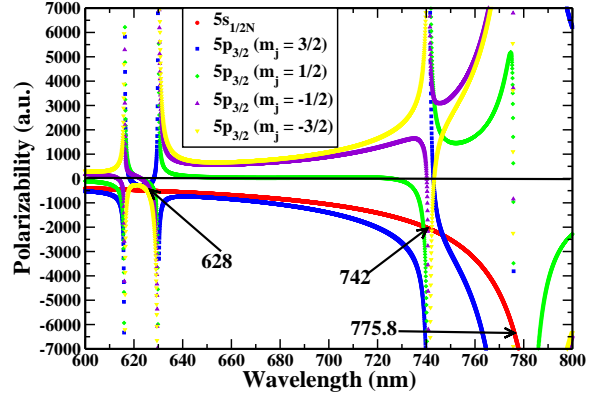}
  \caption{(color online) Magic wavelengths identified by arrows for the $5p_{3/2}-5s$ transition 
in Rb using the left-handed circularly polarized light.}
  \label{figrb-4}     
\end{figure} 

Here we give more emphasis on finding more magic wavelengths
for the $5s-5p_{3/2}$ transition which can be used in the state-insensitive 
trapping scheme for the Rb atom. 
In Table \ref{tabRb-circular2}, we list a number of $\lambda_{\rm{magic}}$ for the $5s-5p_{3/2}$ transition
 in the far-optical and near infrared 
wavelengths along with the uncertainties 
in the $\lambda_{\rm{magic}}$ and the polarizabilities at the $\lambda_{\rm{magic}}$
values. We also list the $\lambda_{\rm{magic}}(\rm{avg})$ values in the table which 
are the average of the magic wavelengths at different $m_j$ sublevels. The error in 
the  $\lambda_{\rm{magic}}(\rm{avg})$ is calculated as the maximum difference between 
the magic wavelengths from different $m_j$ sublevels. 
For this transition we get a set of six magic wavelengths in between seven $5p_{3/2}$ 
resonances lying in the wavelength range 600-1400 $nm$ (i.e. $5p_{3/2}-4d_j$ 
resonance at 1529 $nm$, $5p_{3/2}-6s$ resonance at 1367 $nm$, $5p_{3/2}-5s$ resonance  
at 780 $nm$, $5p_{3/2}-5d_j$ resonance at 776 $nm$, $5p_{3/2}-7s$ resonance at 
741 $nm$, $5p_{3/2}-6d_j$ resonance at 630 $nm$, and $5p_{3/2}-8s$ resonance at 
616 $nm$). Five out of six magic wavelengths support a blue detuned trap (predicted by 
the negative values of dynamic polarizability). Out of these five magic wavelengths the 
magic wavelength at 628 $nm$ and 742 $nm$ are recommended for blue detuned traps.
The magic wavelength at 742 $nm$ supports a stronger trap (as shown by a larger value of the polarizability at this wavelength in Fig.(\ref{figrb-4})). The magic wavelength at 775.8 $nm$ is very close to the resonance and might not be useful for practical purposes. 
The magic wavelength at 1382 $nm$ supports a red detuned optical trap. 
It can be observed from Table \ref{tabRb-circular2} that $m_j=3/2$ sublevel does 
not support state-insensitive trapping at this wavelength. However, using a switching trapping scheme 
as described in the previous paragraph will allow trapping this sublevel too. The magic 
wavelength at 1382 $nm$ is recommended owing to the
fact that  it is not close to any atomic resonance and supports a red-detuned trap which was 
not found in the linearly polarized trapping scheme.

\section{summary}

In conclusion, we have employed the relativistic coupled cluster method
in the singles, doubles and triples excitations approximation to
determine the electric dipole matrix elements in rubidium atom.
Some of the important matrix elements were further optimized 
using the experimental lifetimes of few excited states and static
polarizabilities of the ground and $5p_{1/2,3/2}$ excited states.
These optimized matrix elements were then used to improve the
precision of the available lifetime results for some of the low-lying
excited states in the considered atom. We also observe disagreement
between our calculated dynamic polarizability with a measurement at
the wavelength 1064 $nm$ using the above optimized matrix elements.

We have compared the static and dynamic polarizability results from
various works and reported the improved values of the magic wavelengths 
for the $5s \rightarrow 5p_{1/2}$ transition using the linearly 
polarized light. Issues related to state-insensitive trapping of 
rubidium atoms for the $5s \rightarrow 5p_{3/2}$ transition with
linearly polarized light are discussed and use of the circularly polarized light
is emphasized. Finally, we evaluate six set of magic wavelengths 
for the $5s \rightarrow 5p_{3/2}$ transition which can be used
for the above purpose out of which we have recommended two
magic wavelengths at 628 $nm$ and 742 $nm$ for the blue detuned 
optical traps and 1382 $nm$ for the red detuned optical traps.
We also proposed the use of a switching trapping scheme for the magic 
wavelengths at which the state-insensitive trapping is supported 
only for either positive or negative $m_j$ sublevels of $5p$ states.

\section*{Acknowledgement}
B.K.S. thanks D. Nandy for his help in this work.
The work of B.A. was supported by the Department of Science and
Technology, India. Computations were carried out using 3TFLOP 
HPC Cluster at Physical Research Laboratory, Ahmedabad.

\end{document}